\documentclass{article}
\usepackage{amsmath,amsfonts,amssymb,amsthm}
\usepackage[noend]{algorithmic}
\usepackage{algorithm}
\usepackage{array}
\usepackage{color}
\usepackage{eurosym}
\usepackage[caption=false,font=normalsize,labelfont=sf,textfont=sf]{subfig}
\usepackage{textcomp}
\usepackage{stfloats}
\usepackage{url}
\usepackage{verbatim}
\usepackage{graphicx}
\usepackage{graphics}
\usepackage{cite}

\newcount\cptpropr
\makeatletter
\newenvironment{propri} {
\itshape\global\advance\cptpropr 0\relax
\def\labelitemi{\global\advance\cptpropr\@ne \textsc{\textbf{P\the\cptpropr )}}}
\itemize}{\enditemize}

\newtheorem{thm}{Theorem}
\newtheorem{lemma}{Lemma}
\newtheorem{prop}{Proposition}
\theoremstyle{definition}
\newtheorem{defi}{Definition}
\newtheorem{rem}{Remark}

\begin{document}

\title{Valuing the Electricity Produced Locally in Renewable Energy Communities through Noncooperative Resources Scheduling Games\thanks{This work was partly supported by the Fonds de la Recherche Scientifique - FNRS under grant n°T.0027.21.}} 
\date{}
\author{Louise Sadoine, Zacharie De Grève, and Thomas Brihaye}

%

\maketitle
\begin{abstract}
We propose two market designs for the optimal day-ahead scheduling of energy exchanges within renewable energy communities. The first one implements a cooperative demand side management scheme inside a community where members objectives are coupled through grid tariffs, whereas the second allows in addition the valuation of excess generation in the community and on the retail market. Both designs are formulated as centralized optimization problems first, and as non cooperative games then. In the latter case, the existence and efficiency of the corresponding (Generalized) Nash Equilibria are rigorously studied and proven, and distributed implementations of iterative solution algorithms for finding these equilibria are proposed, with proofs of convergence. The models are tested on a use-case made by 55 members with PV generation, storage and flexible appliances, and compared with a benchmark situation where members act individually (situation without community). We compute the global REC costs and individual bills, inefficiencies of the decentralized models compared to the centralized optima, as well as technical indices such as self-consumption ratio, self-sufficiency ratio, and peak-to-average ratio.

\end{abstract}


\section{Introduction}
T{he} geopolitical crisis which started in Eastern Europe in early 2022 has significantly stressed natural gas markets in the EU, which has in turn driven wholesale electricity prices to unprecedented peaks and volatility in most EU countries \cite{rep-creg}. Retail electricity prices suffered from the same trend: the vast majority of the contract offer in the retail market moved from fixed-price to variable-price contracts, for which the retail price is indexed on wholesale spot markets on a monthly or quarterly basis. Each member state has taken their own measures to protect end-users from the sharp increase in their energy costs, with significant variations between countries, whereas the debate on the relevance of marginal pricing for wholesale electricity markets resurfaced among the policy makers and the scientific community (see e.g. \cite{jour-conejo}).
\\
Renewable Energy Communities (RECs) provide an interesting alternative to partially immune end-users from the high level and volatility of electricity prices. They consist indeed of organized entities, gathering consumers and prosumers connected to the public electricity distribution network, allowed to exchange renewable electricity produced locally without resorting to the wholesale/retail classical structure. Introduced by the EU Commission in its Directive 2018/2001 \cite{Direct}, they aim at 1. placing the citizen at the centre of the liberalized electricity supply chain, by providing (possibly common) economic, environmental or social benefits to its members, 2. stimulating local joint investment in renewable generation and storage assets, and 3. unlocking flexibility inherently present in Low and Medium Voltage (LV and MV) distribution networks.
\\
The mechanisms for exchanging electricity in a community (also called local markets or local communities by some authors) have been extensively studied in the literature, as summarized in \cite{TGP}, \cite{Call22} (and references therein).  For instance, peer-to-peer exchanges within communities of end-users are discussed in \cite{LCJWA19}, the establishment of local demand-supply markets in communities is studied in \cite{CSPGV19,MP18}, whereas cooperative communities built on demand-side management (DSM) schemes are presented in \cite{HTDGV,HTADGV,SHDGB22}. Our work focuses on the latter category, and study extensively the problem of short-term (e.g. day-ahead) dispatch of energy assets within such communities. We argue that a centralized optimization formulation may not be sufficient depending on the local market design, and that the problem requires to resort to game theory in order to model the impact of community members strategies (e.g. schedule of appliances) on the members respective objectives (e.g. electricity bill minimization) and feasible sets.
\\
Mathematical game theory has been considerably investigated in the Smart Grids literature \cite{SHPB12}, and more particularly in the framework of DSM modeling at a community or microgrid scale.
Noncooperative games provide a convenient framework to model interactions between selfish users sharing common network.
Mohsenian-Rad \textit{et al} \cite{MWJSL10} formulated for instance day-ahead energy consumption scheduling games in which each consumer optimizes its own cost by acting on demand scheduling, using a daily billing approach. Atzeni et al. proposed on the other hand a DSM scheme consisting in a day-ahead optimization considering both distributed electricity generation and distributed storage as decision variables rather than shifting energy consumption. The problem has been expressed as noncooperative and cooperative games successively, with a continuous billing strategy \cite{AOSPF13b} and has been extended in \cite{AOSPF14} by embedding global constraints on aggregate bid energy load, thereby creating dependencies between players feasible strategies, leading to a generalized Nash Equilibrium Problem (GNEP). Authors of \cite{MMP22} jointly considered flexible appliances, storage and local dispatchable generation in the DSM process. They solved the game and measured the impact of DSM on system performance parameters. References above do not model however in a community framework which is compliant with EU retail tariff regulation (which separates commodity costs from grid costs), and in which economic flows may be also optimized. Hupez at al. proposed a DSM scheme for communities established on typical European LV grids, and focused on the sharing of costs within the community using game-theoretical billings \cite{HTDGV}, whereas the same authors considered in \cite{HTADGV} a collaborative community where surpluses of local renewable generation and excess storage space are made available freely among community members. The latter hypothesis is however difficultly justifiable in real communities, in which members who have invested in generation assets may wish to value internally (in the community) and externally (through classical markets) their excess production.

We develop in this work two internal market designs for RECs which dictate the exchanges inside the RECs with shiftable appliances, local renewable generation and energy storage systems. The first design (D1) implements a cooperative demand side management scheme inside a community where members objectives are coupled through grid tariffs, whereas the second design (D2) allows in addition the mutualization of excess generation among community members. The contributions of this paper are 3-fold:
\begin{enumerate}
    \item We extend the formalism of \cite{HTADGV} and \cite{HTDGV} by valuing the electricity exchanged internally and sold on the retail market at non-zero prices, and we augment the grid cost structure by considering peak tariffs. We formulate the mathematical problem in a centralized fashion (i.e. optimization-based), and distribute the REC total costs among community members ex-post.
    \item For both market designs, we develop decentralized models building on non cooperative game theory, and endogenize the three cost-distribution approaches in the model. In the first market design (i.e. cooperative demand side management scheme coupled by grid tariffs, D1), we resort to potential game (PG) and variational inequality (VI) theories to prove the existence and study the efficiency of Nash equilibria (NEs). We solve the resulting game in a distributed fashion using the proximal decomposition algorithm \cite{SFPP14}. In the second market design (i.e. which further includes mutualization of excess local generation between REC members, D2), we study the existence and efficiency of Variational Equilibria (VEs), a special class of generalized Nash equilibria (GNEs). We solve the model in a distributed fashion using the proximal algorithm with shared constraints and show convergence under sufficient conditions \cite{AOSPF14}.
    \item We apply the models on a use-case composed of 55 domestic users and compute various community-level (e.g. REC total daily bill, self-consumption, self-sufficiency, etc.) and individual level (member bills) key performance indicators. We compare the obtained results with a benchmark in which the end-users optimize their operation individually, without any community operation. We compare the outcomes of the centralized and decentralized formulations, for each cost distribution. 
\end{enumerate}

The paper is structured as follows. Section \ref{ComFram} describes the community framework, the prosumer load model and the adopted cost structure. Section \ref{Modelisation} presents the day-ahead energy resources scheduling problem for the designs D1 and D2, in both centralized and decentralized cases. We analyze in Section \ref{AnRes} the socially optimal solutions and Nash equilibria. We propose in the same section distributed algorithms for solving the games, and study their convergence. We solve the models on the proposed use-case in Section \ref{Casestu}, and discuss the results, before concluding in Section \ref{concl}.

\section{Community Framework}\label{ComFram}
We assume collaborative communities of consumers and prosumers connected to the same LV public distribution feeder, in which members may virtually mutualize their excess production (D2, see Fig. \ref{fig-1}) or not (D1). Each member is equipped with a bi-directional metering device, or smart meter.

\subsection{Prosumer Profile}
 Let $\mathcal{N} = \{1, \ldots, N\}$ be the set of community members, and $\mathcal{T} = \{1, \ldots, T\}$ the set of time steps of duration $\Delta t$ within a day. The consumption profile of member $i \in \mathcal{N}$ divides into flexible or non-flexible (base) load.

\begin{figure}[t]
\centering
\includegraphics[width=0.9\textwidth]{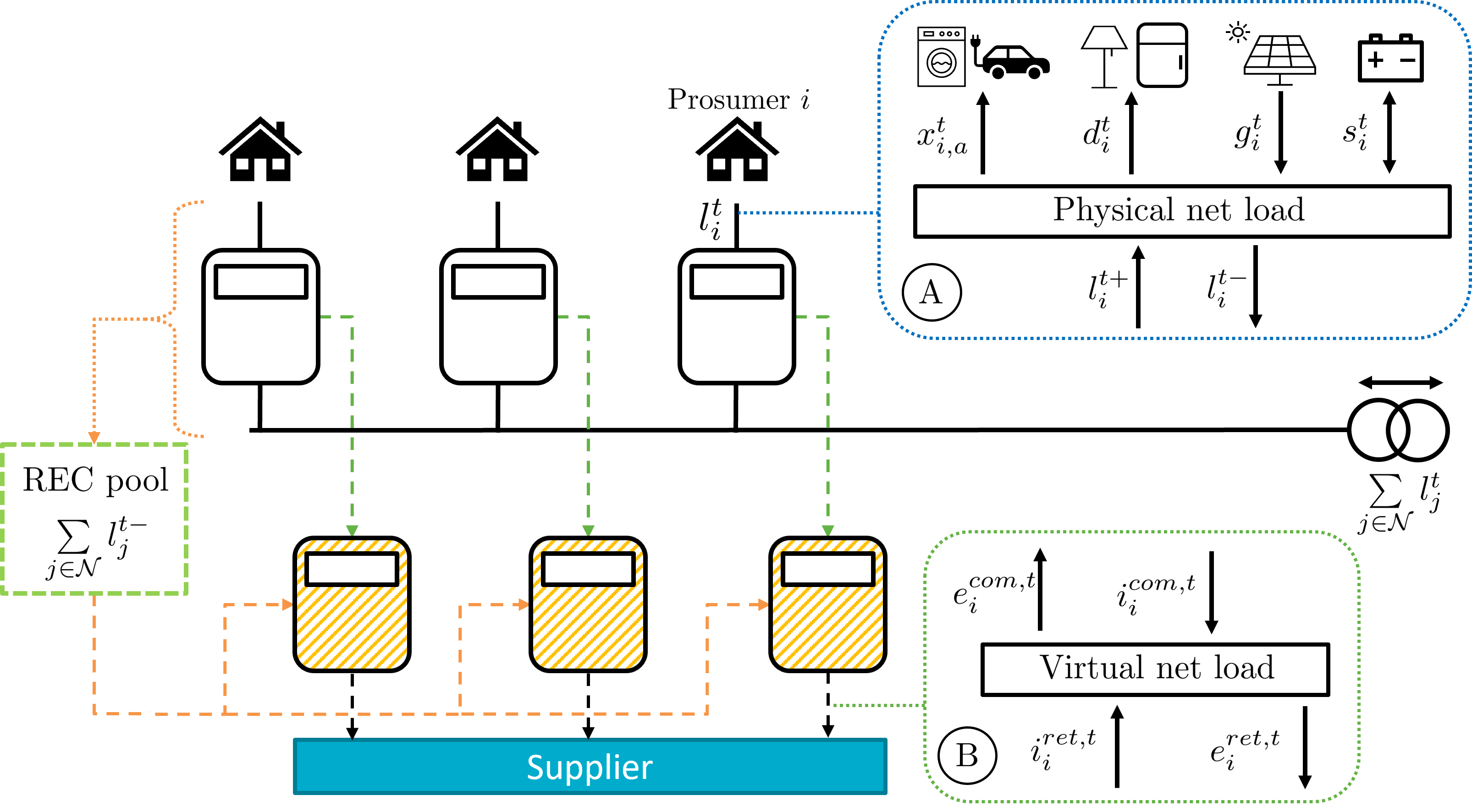}
\caption{Community with mutualization of excess generation (design D2,\cite{SHDGB22})}\label{fig-1}
\end{figure}
 
Let $A_i$ be the set of flexible appliances of member $i$. For each device $a\in A_i$, we define the energy scheduling vector $x_{i,a} = (x_{i,a}^1, \ldots, x_{i,a}^T)$. The non-flexible loads of user $i$ are modeled by $d_i= (d^1_i, \ldots, d^T_i)$ such as $d^t_i \geqslant0, \ \forall t\in\mathcal{T}$. A user $i$ might also be equipped with non-dispatchable energy generation (e.g., photovoltaic panels), represented by $g_i = (g_i^1, \ldots, g_i^T)$ with $g^t_i \geqslant0, \ \forall t\in\mathcal{T}$.
 
Each agent may have a personal energy storage system such as a battery. The power scheduling is given by the storage vector $s_i = (s_i^1, \ldots, s_i^T)$, with $s^t_i > 0$ in charging mode and $s^t_i<0$ discharging mode.

The net load of prosumer $i$ at time $t\in \mathcal{T}$ reads
\begin{equation}
l^t_i= \sum_{a\in A_i} x_{i,a}^t + d^t_i + s^t_i.\Delta t- g^t_i. \label{l1}
\end{equation}
Net load is negative if the prosumer's production exceeds her consumption (export situation), and positive if her local generation does not cover her own consumption (import situation). We define $l^{t+}_i = \max (0,l^t_i)$ and $l^{t-}_i = \max (0, -l^t_i)$, respectively the positive and negative net load, such as $l^t_i = l^{t+}_i - l^{t-}_i$.

The daily peak power consumption of member $i$ reads
\begin{equation*}
    \max_{t\in\mathcal{T}} \left(\dfrac{l^{t+}_i}{\Delta t}\right).
\end{equation*}
We avoid the max operator in the objective function by introducing auxiliary variables $\overline{p}_i$, and reformulate as follows:
\begin{align}
\frac{l^{t+}_i}{\Delta t} &\leqslant \overline{p}_i, \ \ \forall t\in\mathcal{T}. \label{p1}
\end{align}

Flexible appliances are subject to constraints. The temporal flexibility consented to device $a$ by individual $i$ is defined by a daily (parameter) binary vector $\delta_{i,a} = (\delta_{i,a}^1, \ldots, \delta_{i,a}^T)$. A value of 1 indicates that member $i$ agrees to schedule $a$ over time slot $t\in\mathcal{T}$. The predetermined total amount of energy that appliance $a$ must consume for the day is denoted $E_{i,a}$. Without loss of generality, we consider flexible devices with fully modular consumption cycles, i.e., each of them is limited only by maximum power $M_{i,a}$, which reads
\begin{align}
&\delta_{i,a} . x^{\top} _{i,a}= E_{i,a} \label{c1} \\ 
& 0 \leqslant x_{i,a}^t \leqslant M_{i,a}.\delta^t_{i,a} .\Delta t, \ \ \forall t\in \mathcal{T}. \label{c2}
\end{align}

We adopt a simplified storage model neglecting all losses. The battery is subject to maximum charge $M_i^{\text{ch}}$ and discharge $M_i^{\text{dis}}$ power levels, which yields 
\begin{equation}
-M_i^{\text{dis}} \leqslant s_i^t \leqslant M_i^{\text{ch}}  \ \ \forall t\in \mathcal{T}.
\end{equation}
The initial state of charge $e_i^0$ is expressed as a percentage of battery capacity. We impose that the final state of charge is equal to this value. The storage capacity is noted $E^{st}_i$.
\begin{align}
0 \leqslant e_i^0 + \sum_{h=1}^t s^h_i.\Delta t &\leqslant E^{st}_i, \ \ \forall t\in\mathcal{T}\\
e_i^0 + \sum_{t\in\mathcal{T}}s^t_i.\Delta t &= e_i^0.
\end{align}
We consider maximum injection and withdrawal connection power for each member
\begin{align}
l^{t+}_i &\leqslant l_i^{max}, \ \ \forall t\in \mathcal{T}\\
l^{t-}_i &\leqslant g^t_i, \ \ \forall t\in \mathcal{T}\\
\overline{p}_i & \leqslant \frac{l^{max}_i}{\Delta t},\label{lmin}
\end{align}
where $l^{max}_i > 0$ is the upper bound of the member's capacities.\\

In the case of design D2, which allows the virtual mutualization of excess resources among community members (see Fig. \ref{fig-1}), we further define virtual power flows that deviate from the physical flows. A prosumer $i$ with production surplus may sell a quantity $e^{com,t}_i$ at time step $t$  to the community
\begin{equation}\label{sha1}
0\leqslant e^{com,t}_i \leqslant l^{t-}_i, \ \ \forall t\in \mathcal{T}.
\end{equation}
A member $i$ in an energy deficiency situation may also purchase energy $i^{com,t}_i$ from the community
\begin{equation}
0 \leqslant i^{com,t}_i \leqslant l^{t+}_i, \ \ \forall t\in \mathcal{T}.
\end{equation}

Moreover, the total excess production allocated to the community must equal the total quantity imported by members:
\begin{equation}\label{sh2}
\sum_{i\in\mathcal{N}} i^{com,t}_i = \sum_{i\in\mathcal{N}} e^{com,t}_i, \ \ \forall t\in \mathcal{T}.
\end{equation}

Finally, the volumes imported $i^{ret,t}_i$ from and exported $e^{ret,t}_i$ to the retail market by member $i$ are obtained by:
\begin{align}
&i^{ret,t}_i = l^{t+}_i - i^{com,t}, \ \ \forall t\in \mathcal{T} \\
&e^{ret,t}_i = l^{t-}_i - e^{com,t}_i, \ \ \forall t\in \mathcal{T}. \label{eda}
\end{align}

\subsection{Cost Structure}
We assume that members aim at minimizing their energy bill, which is constituted by different components:
\begin{itemize}
\item \textbf{Gray energy costs (D1, D2)}: costs charged by the electricity supplier for the energy not produced locally. We assume a single supplier for the whole community, without loss of generality. The costs are formulated as $C_{ret}(l^{t+}_i)= \lambda_{imp}^t.l^{t+}_i$, with $\lambda_{imp}^t$ in \euro /kWh.

\item \textbf{Local energy costs (D2 only):} costs of electricity bought from the REC pool, at tariff $\lambda_{iloc}^t$ \euro /kWh. For each user $i\in\mathcal{N}$, we have $C_{loc}(i_i^{com,t})=\lambda_{iloc}^t.i_i^{com,t}$.
\item \textbf{Revenue from exported energy (D1, D2):} income related to the sale of excess local production on the retail market at $\lambda_{exp}^t$ \euro /kWh, and to the local community at  $\lambda_{eloc}^t$ \euro /kWh (D2 model). We have: $R_{ret}(l^{t-}_i) = \lambda_{exp}^t.l^{t-}_i$ and $R_{loc}(e^{com,t}_i)= \lambda_{eloc}^t. e^{com,t}_i $ (D2 only).
\item \textbf{Grid costs (D1, D2):} we assume that grid costs are divided into two components. \textbf{Upstream Grid Costs} account for the use of the upstream MV and HV grids, and are expressed as $C_{gr}^t= \alpha.(L^t)^2 = \alpha.( \sum_{i\in\mathcal{N}} l^t_i )^2$, with $L^t=\sum_{i\in\mathcal{N}}l^t_i$ the aggregated net load at the MV/LV transformer, and with  
with $\alpha$ \euro /kWh$^2$ the price for the upstream network use. \textbf{Local Grid Costs} are assumed to be purely capacity-based, and are computed based on the daily peak consumption of each member as $C_{i,peak}$ = $\beta.\overline{p}_i$, with $\beta$ \euro /kW unit penalty cost.
\end{itemize}

The total REC costs for D1 and D2 are respectively:
\begin{align}
f^1(\Theta) =& \sum\limits_{t\in\mathcal{T}} \big[\sum \limits_{i\in\mathcal{N}} (C_{ret}(l^{t+}_i)-R_{ret}(l^{t-}_i)) + C_{gr}^t\big] + \sum\limits_{i\in\mathcal{N}} \beta.\overline{p}_i \label{f1} \\
f^2(\Theta) =& \sum\limits_{t\in\mathcal{T}} \big[\sum \limits_{i\in\mathcal{N}} (C_{ret}(i^{ret,t}_i) + C_{loc}(i^{com,t}_i)-R_{ret}(e^{ret,t}_i) \notag\\
&-R_{loc}(e^{com,t}_i)) + C_{gr}^t\big] + \sum\limits_{i\in\mathcal{N}} \beta.\overline{p}_i \label{f2}
\end{align}
$\Theta_i$ is the set of decision variables of member $i$, with $\Theta := (\Theta_1,\ldots, \Theta_N)$. We assume $\lambda_{exp}^t < \lambda_{imp}^t$ and $\lambda_{eloc}^t<\lambda_{iloc}^t$.

\section{Day-Ahead Energy Resources Scheduling Problem}\label{Modelisation}
We formulate the day-ahead energy scheduling problem, in which members optimize their available flexibility (D1 and D2) and the virtual energy exchanges (D2) so as to minimize the total energy costs. Section \ref{optiproblem} presents the centralized (i.e. optimization-based) formulation, whereas Section \ref{GT} describes the decentralized (i.e. game theoretical) formulation. 

\subsection{Centralized optimization models}\label{optiproblem}
\textbf{Design D1} consists in a cooperative demand-side management scheme between community members, coupled via  the upstream grid cost component.  We assume a central operator (e.g. a community-manager) is solving the model in order to minimize the total REC electricity costs
 \begin{equation}\label{CNM}
 \begin{split}
\underset{\Theta}{\min}  &\ f(\Theta) \text{ as in (\ref{f1})} \\
\text{s.t. }  &\Theta \in \Omega^1
\end{split}
\end{equation}
with $\Omega^1:= \{(x_i,s_i,l_i^+,l_i^-,\overline{p}_i)_{i=1}^N \in \mathbb{R}^n: (\ref{l1})- (\ref{lmin})\}$ the feasible set. 

\textbf{Design D2} allows prosumers to sell their excess production to the community pool, whereas consumers may purchase electricity on that pool. We have
\begin{equation}\label{CM}
\begin{split}
\underset{\Theta}{\min} &\ f(\Theta) \text{ as in (\ref{f2})}  \\
\text{s.t. }  &\Theta \in \Omega^2
\end{split}
\end{equation}
where $\Omega^2:= \{(x_i, s_i, i_i^{com}, e_i^{com}, i_i^{ret}, e_i^{ret}, l_i^+,l_i^-,\overline{p}_i)_{i=1}^N \in \mathbb{R}^n: (\ref{l1})- (\ref{eda})\}$ is the feasible set. The profiles minimizing the total bill are named \textit{socially optimal solutions}.

\subsection{Noncooperative Games}\label{GT}
Designs D1 and D2 may give rise to strategic interactions between the community members, who compete for common resources (i.e. the  network and the local production surplus), which are not captured by (\ref{CNM})-(\ref{CM}). Additionally, the allocation of costs to each member is not addressed in the centralized models, which minimize the total REC electricity bill only (cost allocation must be performed ex-post in that case). We formulate consequently the day-ahead energy resources scheduling problem as a non-cooperative game for D1 and D2.
We assume three mechanisms for allocating REC costs among individuals and compute the individual bills $b_i(\Theta)$, inspired by \cite{HTADGV}. The two first share the total REC daily costs $f(\Theta)$ proportionally among members according to distribution keys $K_i$, whereas the third one allocates the total REC costs on a hourly basis. Note that other allocation mechanisms, based on energy sharing rather than cost sharing, can be found in the literature (see e.g. \cite{MRMDP22}).

\begin{enumerate}
\item \textbf{Net load proportional billing [Net].} The distribution key for member $i$ is given by the ratio between the absolute value of her net load and the community net load. More precisely, we consider the minimal value of the net load, i.e. $l^{t*}_i=\min(\sum_{t\in\mathcal{T}}|l^t_i|)$, in order to make the key independent of the solution set \cite{HTADGV}.
    \begin{equation}\label{net}
b^{\text{Net}}_i(\Theta)=K_i f(\Theta)=\frac{\sum_{t\in\mathcal{T}}l^{t*}_i}{\sum_{j\in\mathcal{N}}\sum_{t\in\mathcal{T}}l^{t*}_j} f(\Theta)
\end{equation}
\item \textbf{Marginal cost billing [VCG].}
The distribution key of each member is calculated using the normalized Vickrey-Clarke-Groves mechanism \cite{JBGO17, HTADGV}
\begin{equation}\label{vcg}
b^{\text{VCG}}_i(\Theta)=K_i f(\Theta)=\frac{|C_{\mathcal{N}}^*-C_{\mathcal{N}\backslash\{i\}}^*| }{\sum_{j\in\mathcal{N}}|C_{\mathcal{N}}^*-C_{\mathcal{N}\backslash\{j\}}^*|} f(\Theta)
\end{equation}
where $C_{\mathcal{N}}^*$ are the minimum community costs for a REC composed by all members of set $\mathcal{N}$, and are obtained by solving \eqref{CNM} or \eqref{CM}. This requires the solution of $N$ additional optimization problems.
\item \textbf{Continuous proportional billing [CP].}
The total cost is distributed among community members at each time slot $t\in\mathcal{T}$. We have for designs D1 and D2 respectively
\begin{subequations}
\begin{align}
b_i^{\text{CP1}}(\Theta) =& \sum\limits_{t\in\mathcal{T}} (\lambda_{imp}^t.l^{t+}_i -\lambda_{exp}^t.l^{t-}_i + l^t_i.\alpha L^t) + \beta.\overline{p}_i \label{b1} \\
b_i^{\text{CP2}}(\Theta) =& \sum\limits_{t\in\mathcal{T}} (\lambda_{imp}^t.i^{ret,t}_i + \lambda_{iloc}^t.i^{com,t}_i- \lambda_{eloc}^t.e^{com,t}_i \notag\\
&- \lambda_{exp}^t.e^{ret,t}_i  + l_i^t.\alpha L^t) + \beta.\overline{p}_i \label{b2}
\end{align}
\end{subequations}
where $L^t=\sum_{j\in\mathcal{N}}l^t_j$.\\
\end{enumerate}

\subsubsection{Design D1 as a NEP} \label{subsubsec-NEPmodel} we rewrite problem (\ref{CNM}) as a Nash equilibrium problem (NEP), in which each member $i\in\mathcal{N}$ (or agent) is a selfish player choosing her strategy profile $\Theta_i$ in order to minimize her own bill $b_i : \mathbb{R}^{n} \to \mathbb{R}$, which depends itself on other players' strategies $\Theta_{-i}:=(\Theta_1,\ldots,\Theta_{i-1}, \Theta_{i+1},\ldots,\Theta_{N})$. Mathematically, agent $i$ solves the following optimization problem
\begin{equation}
G:=
    \begin{cases} \label{NEP}
         \underset{\Theta_i}{\min} & b_i(\Theta_i, \Theta_{-i}) \ \ \ \ \ \forall i\in\mathcal{N} \\
         \text{s.t. } & \Theta_i\in \Omega_i
    \end{cases}
\end{equation}
where $\Omega_i \subseteq \mathbb{R}^{n_i}$ is the strategy set constituted by the player $i$'s individual constraints, which are independent of the other members strategies in D1. The $n$-dimensional joint strategy set is expressed as $\Omega := \prod_{i\in\mathcal{N}}\Omega_i$, with $n :=\sum_{i\in\mathcal{N}}n_i$. The game $G$ is described as the triplet $G=(\mathcal{N},\Omega,(b_i)_{i=1} ^N)$.

A strategy profile $\Theta^*\in\Omega$ is called a Nash equilibrium (NE) of the game $G$ (\ref{NEP}) if $\forall i\in\mathcal{N}$:
\begin{equation}\label{NE}
    b_i(\Theta^*_i,\Theta^*_{-i}) \leqslant b_i(\Theta_i, \Theta^*_{-i}), \ \ \forall \Theta_i\in\Omega_i.
\end{equation}
The set of Nash equilibria of game $G$ is denoted NE$(G)$. A NE is a feasible strategy profile such that no single player can benefit by unilaterally deviating from her strategy. Note that a NE does not necessarily lead to the social optimum, and the social optimum may not be a NE.\\

\subsubsection{Design D2 as a GNEP}\label{subsubsec-GNEPmodel} design D2, which allows mutualization of excess production among community members, comes with additional constraints ensuring the balance of internal virtual exchanges (\ref{sh2}). These constraints couple the strategy set of each player to her rivals' decisions. We reformulate consequently problem (\ref{CM}) as a Generalized Nash Equilibrium Problem (GNEP), in which both the objective functions and the strategy sets depend on the rivals' strategies, contrarily to NEPs where interactions occur on the objective functions only. We have
\begin{equation}
\mathcal{G}:=
    \begin{cases} \label{GNEP}
         \underset{\Theta_i}{\min} &  b_i(\Theta_i, \Theta_{-i}) \ \ \ \ \ \forall i\in\mathcal{N} \\
         \text{s.t. } & \Theta_i\in \Omega_i(\Theta_{-i})
    \end{cases}
\end{equation}
where the strategies of player $i$ must belong to her feasible strategy set  $\Omega_i(\Theta_{-i})\subseteq\mathbb{R}^{n_i}$. Similarly, the game is described as the triplet $\mathcal{G}=(\mathcal{N},(\Omega_i(\Theta_{-i}))_{i\in\mathcal{N}},(b_i)_{i\in\mathcal{N}})$.

In our case, shared constraints are linear so that we define the feasible set of agent $i$ as
\begin{equation}
    \Omega_i(\Theta_{-i}):= \{\Theta_i \in\overline{\Omega}_i : h(\Theta_i,\Theta_{-i})=0 \},
\end{equation}
where $\overline{\Omega}_i\subseteq \mathbb{R}^{n_i}$ is player $i$ individual constraints set. We write the shared coupling constraints (\ref{sh2}) as: $h(\Theta) := (\sum_{i\in\mathcal{N}}e^{com,t}_i-i^{com,t}_i)_{t\in\mathcal{T}}$. The joint strategy set reads
\begin{equation}\label{jointcon}
    \mathcal{C} = \{\Theta\in\mathbb{R}^n : \Theta_i\in\overline{\Omega}_i \ \forall i\in\mathcal{N}, \ h(\Theta)=0\}.
\end{equation}

A strategy profile $\Theta^*$ is called a Generalized Nash Equilibrium of the game $\mathcal{G}$ (\ref{GNEP}) if $\forall i\in \mathcal{N}:$
\begin{equation}
    b_i(\Theta^*_i, \Theta^*_{-i}) \leqslant b_i(\Theta_i,\Theta^*_{-i}), \ \ \forall \Theta_i\in\Omega_i(\Theta^*_{-i}).
\end{equation}
The set of generalized Nash equilibria of $\mathcal{G}$ reads GNE$(\mathcal{G})$.

\section{Analysis and Resolution}\label{AnRes}
This section studies the existence of solutions and presents solution algorithms for each model presented in Section \ref{Modelisation}.
\subsection{Centralized Optimization Problems}
We first analyze the classification and properties of centralized optimization problems \eqref{CNM} and \eqref{CM}. 
\begin{thm}
The optimization problems \eqref{CNM} and \eqref{CM} have at least a global minimum.
\end{thm}

The proof is straightforward. The cost functions (\ref{f1}) and (\ref{f2})) are quadratic, continuously differentiable and convex (we can easily show that their Hessian matrix is positive semi-definite). Furthermore, each feasible set is a bounded polyhedron, and is assumed non empty. Functions $f$ are therefore bounded on the feasible set and reach their bounds.  
These models are therefore convex quadratic optimization problems, for which the existence of a solution is guaranteed (multiple solutions may exist however, since the objective functions $f$ are not strongly convex).
These problems can be solved in a centralized way by standard algorithms such as interior-point methods, in polynomial time \cite{BV04, KTK80}.

\subsection{Nash Equilibrium Problems}
We study the existence of Nash equilibria in the case of design D1 (section \ref{subsubsec-NEPmodel}), for each cost distribution method, and propose a solution algorithm that converges to a NE under sufficient conditions \cite{SFPP14}.

\subsubsection{Nash Equilibrium Analysis}\label{NEana}

\begin{thm}\label{exine} Given the NEP \eqref{NEP}, for the three cost distribution methods:
\begin{enumerate}
    \item The game possesses at least one Nash equilibrium: NE($G$)$\neq \emptyset$,
    \item The NE set is compact.
\end{enumerate}
\end{thm}

\begin{thm}\label{optine} Given the NEP \eqref{NEP}
\begin{enumerate}
    \item For [Net] and [VCG], the game is equivalent to the optimization problem \eqref{CNM}, and all the NEs $\Theta^*$ of $G$ lead to the same optimal values of the individual objective functions and $f$ \eqref{f1}.
    \item For [CP], the game is equivalent to the optimization problem: 
    \begin{equation}\label{potcont}
    \begin{split}
    \underset{\Theta}{\min} & \ f^1(\Theta) - \frac{\alpha}{2}\sum_{t\in\mathcal{T}}\sum_{i\in\mathcal{N}}l^t_i.L^t_{-i}\\
    \text{s.t. } & \Theta\in\Omega
    \end{split}
    \end{equation}
    where $L^t_{-i} = \sum_{j\in\mathcal{N}\backslash\{i\}}l^t_j.$ The NEs may deviate from the social optimum.
\end{enumerate}
\end{thm}

Theorem \ref{exine} guarantees the existence of at least one NE for game $G$. Theorem \ref{optine} states that NEP \eqref{NEP} can be formulated as a centralized optimization problem. More particularly, for [Net] and [VCG] cost distribution methods, the set of NEs coincide with the optimal solutions of the social problem \eqref{CNM}. The optimal values of players' billing functions are constant over the NE set. Roughly speaking, one can say that all the NEs are equivalent (in terms of optimal values of the players' cost functions)

We prove both theorems by relying on specific properties of our problem, and by resorting to results from Potential Game theory and Variational Inequality theory.  It can first easily be shown that the following properties hold

\begin{propri}
    \item Each $b_i$ is continuously differentiable on $\Omega$.
    \item Each $\Omega_i$ is closed and convex.
    \item Each $\Omega_i$ is bounded.
    \item For any $\Theta_{-i}$, $b_i(\cdot, \Theta_{-i})$ is convex on $\Omega_i$.
\end{propri}

We then show that NEP \eqref{NEP} is a Potential Game. A game $G=(\mathcal{N},\Omega,(b_i)_{i=1} ^N)$ is a potential game if there exists a single global function, or potential function, $P: \Omega \to \mathbb{R}$ describing the consequences of a change in strategy for each player \cite{MS96}. The game $G$ is a \textit{weighted potential game} (WPG) if for each member $i$ and $\Theta_i,\Theta'_i\in\Omega_i$, $\Theta_{-i}\in\Omega_{-i}$, $b_i(\Theta'_i,\Theta_{-i}) - b_i(\Theta_i, \Theta_{-i}) = \omega_i (P(\Theta'_i,\Theta_{-i}) - P(\Theta_i,\Theta_{-i}))$, with $(\omega_i)_{i\in\mathcal{N}}$ a vector of positive numbers. When $\omega_i=1$ for every $i\in\mathcal{N}$, the game $G$ is an \textit{exact potential game} (PG). For the remainder of the paper, we name a potential game any game with a weighted or exact potential function.\\

The NEP \eqref{NEP} with [Net] and [VCG] is a WPG, with $P(\Theta)=f^1(\Theta)$ in \eqref{f1} and $\omega=(K_i)_{i\in\mathcal{N}}$. Each member's individual bill is given by the total energy costs times a strictly positive constant $K_i$. Therefore, all the NEs lead to the same payoff function value. The NEP \eqref{NEP} with [CP] is a PG, with $P(\Theta)=f^1(\Theta)- (\alpha/2).\sum_{t\in\mathcal{T}}\sum_{i\in\mathcal{N}}l^t_i.L^t_{-i}$. It can easily be shown by definition of potential game and \eqref{NE}, that an optimal solution of \eqref{CNM}/\eqref{potcont} is a NE. Conversely, as the potential functions are continuously differentiable and convex on $\Omega$, a NE is an optimal solution of \eqref{CNM}/\eqref{potcont} \cite{SBP06}, showing Theorem \ref{optine}. The characteristics of our NEP can be obtained by solving standard optimization problems. Furthermore, given properties \textbf{P2} and \textbf{P3}, we show that the NE set is non-empty by applying \cite[Lem. 4.3]{MS96}, proving Theorem \ref{exine}, 1).

We further resort to the theory of finite-dimensional Variational inequalities (VIs), in order to prove the second point of Theorem \ref{exine}, and to benefit from powerful solution algorithms available in the literature \cite{FP07}.

\begin{defi}[Variational Inequality \cite{FP07}] Consider a set $\mathcal{Y}\subseteq\mathbb{R}^n$ and a mapping $F: \mathcal{Y} \to \mathbb{R}^n$. A solution to the variational inequality problem VI$(\mathcal{Y},F)$, is a vector $y^*\in\mathcal{Y}$ such that
\begin{equation}
    (y-y^*)^\top F(y^*) \geqslant 0, \ \ \forall y\in\mathcal{Y}.
\end{equation}
\end{defi}

The set of solutions is denoted SOL$(\mathcal{Y},F)$. For each player $i\in\mathcal{N}$, the strategy set $\Theta_i$ verifies \textbf{P2} and is assumed nonempty. The payoff functions $b_i$ verify \textbf{P1} and \textbf{P4}. Hence, the NEP is equivalent to the VI formulation \cite[Prop. 1.4.2]{FP07}.
\begin{prop}\label{existvi}
The NEP $G=(\mathcal{N},\Omega,(b_i)_{i=1}^N)$ is equivalent to the VI$(\Omega,F)$, with $F(x):=(\nabla_{\Theta_i} b_i(\Theta))_{i=1}^N$. We have NE$(G)$ = SOL$(\Omega, F)$.
\end{prop}

Since each $\Omega_i$ verifies \textbf{P3}, hence so is $\Omega$. Thus, by \cite[Prop. 2.2.9]{FP07},  Theorem \ref{exine} is also established.

\subsubsection{Nash Equilibrium Computation}\label{NEcomp} centralization raises important issues related to prosumers' consumption privacy. We focus on the computation of the NEs of game \eqref{NEP}, via distributed iterative algorithms. Since the problem may have multiple NEs, the classical best-response algorithm  may fail to converge~\cite{SFPP14}. Hence, we adopt a proximal decomposition algorithm (PDA) that have desirable privacy-preserving properties \cite{AOSPF13b},\cite{SFPP14}. Instead of a single NEP, we solve a sequence of strongly convex sub-problems with a particular structure which are guaranteed to converge under some technical conditions. We consider a regularization of the original VI$(\Omega,F)$, given by VI$(\Omega, F+\tau(I-y^k))$, where $I$ is the identity map, $y^k$ is a fixed vector in $\mathbb{R}^n$ and $\tau$ is a positive constant. At each iteration $k+1$, the players update their strategies simultaneously (via a Jacobi scheme) by minimizing their bill while perceiving the recently available value of the aggregate net load. The regularized VI is therefore equivalent to the following game
\begin{equation}
    \begin{split}
    \underset{\Theta_i}{\min} & \ b_i(\Theta_i,\Theta_{-i}^k)+ \frac{\tau}{2} || \Theta_i -\Theta_i^k||^2\\
    \text{s.t. } & \Theta_i\in\Omega_i.
    \end{split}
\end{equation}

\begin{algorithm}[H]
\caption{Proximal Decomposition Algorithm (PDA)}

\begin{algorithmic}
\STATE Choose any starting point $\Theta^0\in\Omega$, set $k=0$.
   Given $\{\rho^k\}_{k=0}^\infty$ and $\tau >0.$
\WHILE{\text{a suitable termination criterion is not satisfied}}
\FOR{$i\in\mathcal{N}$, each member computes $\Theta^{k+1}_i$ as}
\STATE $\Theta_i^{k+1} := \Theta^*\in \arg\min\{ b_i(\Theta_i,\Theta^{k}_{-i})$ + $\frac{\tau}{2}||\Theta_i - \Theta_i^k||^2, \Theta_i\in\Omega_i \}$
\ENDFOR
\IF{the NE is reached} 
\STATE each player $i\in\mathcal{N}$ sets $\Theta^{k+1}_i \gets (1-\rho^k).\Theta^k_i + \rho^k.\Theta_i^{k+1}$
\ENDIF
\STATE $k\gets k+1$
\ENDWHILE
\end{algorithmic}
\end{algorithm}

Convergence of Algorithm 1 is studied in Theorem \ref{algonep}.
\begin{thm}\label{algonep} Let the game $G$ \eqref{NEP} be a NEP with a nonempty set solution. If 
\begin{enumerate}
    \item the regularization parameter $\tau$ satisfies for [Net,VCG]:
    \begin{subequations}
        \begin{align}
        &\tau > 4\alpha (N-1) \max_{i\in\mathcal{N}}K_i \\
        \intertext{and for [CP]:}
        &\tau > 2\alpha (N-1)
        \end{align}
    \end{subequations}
    \item $\rho$ is chosen such that $\rho \subset [R_m, R_M]$, with $0< R_m < R_M < 2$,
\end{enumerate}
then, any sequence $\{\Theta^k\}_{k=1}^{\infty}$ generated by PDA converges to a Nash equilibrium of the game.
\end{thm}

The proof is straightforward in the case of the [CP] billing, since the monotonicity of $F$ ensures directly convergence \cite{SFPP14}. The proof of [Net,VCG] involves Theorem \ref{optine}.1) and \textbf{P3}. See Appendix \ref{appendixA}.

\subsection{Generalized Nash Equilibrium Problems}
GNE problems are difficult to solve because the strategy sets depend on the rivals strategies.
They can usually be expressed as quasi-variational inequality (QVI) problems \cite{FP07}. Despite recent progresses \cite{K16,KS18}, solution algorithms are computationally intensive, and distributed implementations remain an open problem \cite{F12}. We study the existence and unicity of GNEs for GNEP \eqref{GNEP}, and resort to the proximal Jacobi best-response algorithm with shared constraints, which is able to find a GNE under sufficient conditions \cite{SFPP14}.
\subsubsection{Generalized Nash Equilibrium analysis}

\begin{thm}\label{exignep} Let us consider GNEP \eqref{GNEP}, for all cost distribution methods. The game possesses at least one generalized Nash equilibrium: GNE$(\mathcal{G})\neq\emptyset$.
\end{thm}
Theorem \ref{exignep} guarantees the existence of a GNE for game $\mathcal{G}$, but it may have multiple solutions. The proof is described below.

Each payoff function $b_i$ fulfills \textbf{P1} and \textbf{P4}. The local constraint sets  $\overline{\Omega}_i$ verify \textbf{P2} and \textbf{P3}, and are assumed non-empty. We also verify that
\begin{propri}
\item Function $h$ is continuous and componentwise convex.
\end{propri}
The joint strategy set $\mathcal{C}$ \eqref{jointcon} satisfies therefore \textbf{P2} and \textbf{P3}, and the GNEP \eqref{GNEP} is \textit{jointly convex} \cite{FK07}: an equilibrium can be calculated by solving a suitable VI problem.

\begin{prop}[\cite{FK07}]\label{vigne} Let us consider the jointly convex GNEP $\mathcal{G}$ \eqref{GNEP}, and let $F(\Theta):= (\nabla_{\Theta_i}b_i(\Theta_i,\Theta_{-i}))_{i=1}^N.$ Then, every solution of the VI$(\mathcal{C},F)$ is a solution of the GNEP: SOL$(\mathcal{C},F)\subseteq$ GNE$(\mathcal{G})$.
\end{prop}
We do not have however that any solution of the GNEP is also a solution of the associated VI \cite{FK07}. The solutions of the GNEP that are also solutions of the VI$(\mathcal{C},F)$ are called \textit{variational equilibria} (VEs). We then analyze and compute the variational solutions of the original GNEP \eqref{GNEP}. Since the joint strategy set $\mathcal{C}$ holds \textbf{P3}, the result \cite[Prop. 2.2.9]{FP07} guarantee existence of the VE set and thus ensure the existence of the GNE for the game \eqref{GNEP}. Theorem \ref{exignep} is therefore established.

Our GNEP has a special structure that may simplify the equilibrium analysis. We have 

\begin{thm}\label{optignep} Given the jointly convex GNEP \eqref{GNEP}
\begin{enumerate}
    \item For [Net,VCG], the social minimum solutions of \eqref{CM} are included in the GNE set. If the $K_i$'s are equal, the VE set coincides with the social optimum over the set $\mathcal{C}$, and lead to the same value of the objective functions.
    \item For the billing function \eqref{b2}, the VE set coincide with the optimal solution of the optimization problem:
    \begin{equation}\label{potcont2}
     \begin{split}
    \underset{\Theta}{\min} & \ f^2(\Theta) - \frac{\alpha}{2}\sum_{t\in\mathcal{T}}\sum_{i\in\mathcal{N}}l^t_i.L^t_{-i}\\
    \text{s.t. } & \Theta\in\mathcal{C}
    \end{split}
    \end{equation}
    where $L^t_{-i} = \sum_{j\in\mathcal{N}\backslash\{i\}}l^t_j.$ The VEs may not achieve social optimum.
\end{enumerate}
\end{thm}
For [Net,VCG], we prove that all the optimal solutions of the social problem \eqref{CM} are included in the GNE set. For [CP], and for [Net,VCG] under specific conditions, calculating the VE of game \eqref{GNEP} is similar to solving an optimization problem. We prove theorem \ref{optignep} below.

Again, we exploit connections with potential games in the case of GNEPs \cite{FPS11}. The jointly convex GNEP \eqref{GNEP} under [Net,VCG] is a WPG, with $P(\Theta)=f^2(\Theta)$ in \eqref{f2} and $\omega=(K_i)_{i\in\mathcal{N}}$. Also, the GNEP with [CP] is a PG, with $P(\Theta)=f^2(\Theta)-(\alpha/2).\sum_{t\in\mathcal{T}}\sum_{i\in\mathcal{N}}l^t_i.L^t_{-i}$. Then, it can be easily checked that an optimal solution of \eqref{CM}/\eqref{potcont2} is a GNE of the GNEP. Note that a GNE $\Theta^*$ of the GNEP \eqref{GNEP} does not always minimize the potential function over the joint strategy set $\mathcal{C}$ \eqref{jointcon}, due to shared constraints \cite{YRSP13}.
Nevertheless, the continuous billing PG have its variational equilibrium set coinciding with the optimal solutions. In fact, \eqref{potcont2} is a convex optimization problem hence a point $\Theta$ is a minimum  if and onfly if $(\Theta' - \Theta)^\top \nabla_{\Theta}P(\Theta) \geqslant 0$ for all $\Theta'\in\mathcal{C}$  \cite[(4.21)]{BV04}. Because $\nabla_{\Theta}P(\Theta)=F(\Theta)$, the latter coincides with $(\Theta'-\Theta)^\top F(\Theta) \geqslant 0$ for all $\Theta' \in\mathcal{C}$ and so $\Theta\in$ SOL$(\mathcal{C},F)$.

\subsubsection{Generalized Nash Equilibria Computation}
We focus on the distributed computation of the GNEs, especially the VEs. Inspired by the VI framework proposed in \cite{AOSPF14} and \cite{SPFP11}, we apply the PDA with coupling constraints.

Distributed algorithms require the joint strategy set to be a Cartesian product, so that a VE cannot be computed from the VI formulation. We decouple the members' feasible sets by converting the global constraints into penalty terms in the objective functions. A new player $N+1$ is also introduced. The extended NEP reads
\begin{equation}\label{Gext}
\mathcal{G}_{ext}:= 
\begin{cases}
\underset{\Theta_i\in \overline{\Omega}_i}{\min}  \ b_i(\Theta_i,\Theta_{-i}) + \pi^{\top}h(\Theta) \ \ &\forall i\in\mathcal{N}\\
\underset{\pi\in\mathbb{R}^T}{\min}  \ -\pi^{\top}h(\Theta) \ \ \ &i=N+1.
\end{cases}
\end{equation}
This new player can be considered as a central operator (e.g. a community-manager) controlling a price variable $\pi\in\mathbb{R}^T$, that can be interpreted as prices associated with an imbalance between energy exported to and imported from the REC pool, represented by the global constraints $h(\Theta)$.
 
In fact, $\pi$ is the Lagrange multiplier associated with the constraints $h(\Theta)$.
We have the following property
\begin{propri}
    \item The  $\overline{\Omega}_i$'s and the joint strategy set $\mathcal{C}$ satisfy the Slater's constraint qualification \cite[(5.27)]{BV04}.
\end{propri}

The connection between the solutions of the GNEP $\mathcal{G}$ and $\mathcal{G}_{ext}$ is displayed in the Lemma 1 \cite{PGPKL16}.
\begin{lemma}[Extended game] A point $(\Theta ^*, \pi^*) \in\mathcal{C} \times \mathbb{R}^T$ is a Nash equilibrium of the game $\mathcal{G}_{ext}$ if and only if $\Theta^*$ is a solution of the VI$(\mathcal{C},F)$ with multiplier $\pi^*$ associated with the shared constraints $h(\Theta^*)=0$.  
\end{lemma}

We have transformed the computation of VEs of the GNEP \eqref{GNEP} into solving an extended NEP \eqref{Gext}. The NEs of $\mathcal{G}_{ext}$ coincide with the solutions of an extended VI problem that can be achieved via distributed algorithms. 

\begin{lemma}[Extended VI]\label{lem2}
A point $(\Theta ^*, \pi^*) \in\mathcal{C} \times \mathbb{R}^T$ is a Nash equilibrium of the game $\mathcal{G}_{ext}$ if and only if it is a solution of the VI($\mathcal{Y}, F_{ext})$, with $\mathcal{Y}:=(\prod_{i\in\mathcal{N}} \overline{\Omega}_i) \times \mathbb{R}^T$ and
\begin{equation}
    F_{ext}(\Theta,\pi):= 
    \begin{bmatrix}
    (\nabla_{\Theta_i} b_i(\Theta_i,\Theta_{-i}) + \nabla_{\Theta_i}\pi ^\top h(\Theta_i,\Theta_{-i}))_{i=1}^N \\
    -h(\Theta)\\
    \end{bmatrix}
\end{equation}
\end{lemma}
We summarize the previous results with Theorem \ref{extvi}.
\begin{thm}\label{extvi} A point $\Theta^*$ is a variational equilibrium of the GNEP \eqref{GNEP} if and only if a $\pi^*$ exists such that $(\Theta^*,\pi^*)$ is a solution of the extended VI$(\mathcal{Y}, F_{ext})$.
\end{thm}

We have decoupled the members' constraints by incorporating the global constraints $h(\Theta)$ into the objective function. Thanks to the reformulation in Theorem \ref{extvi}, we obtain the set $\mathcal{Y}$ as the Cartesian product of individual feasible sets. Hence, as in the Section \ref{NEcomp}, we solve a regularized sequence of VI$(\mathcal{Y}, F_{ext}+\tau(I-(y^k,\eta^k)))$ with $(y^k,\eta^k)$ in $\mathbb{R}^{n+T}$.
\begin{algorithm}[H]
\caption{PDA with shared constraints}

\begin{algorithmic}
\STATE Choose any starting point $\Theta^0\in\mathcal{Y}$, set $k=0$.
   Given $\{\rho^k\}_{k=0}^\infty$ and $\tau >0.$
\WHILE{\text{a suitable termination criterion is not satisfied}}
\FOR{$i\in\mathcal{N}$, each member computes $\Theta^{k+1}_i$ as}
\STATE $\Theta_i^{k+1} := \Theta_i^* \in \arg\min\{ b_i(\Theta_i,\Theta^{k}_{-i}) + \pi^\top h(\Theta_i,\Theta_{-i}^k)$ + $\frac{\tau}{2}||\Theta_i - \Theta_i^k||^2, \Theta_i\in\Omega_i \}$
\ENDFOR
\STATE The central operator computes $\pi^{k+1}$ as \\
$\pi^* \in\arg\min\{ -\pi^\top h(\Theta^k) + \frac{\tau}{2}||\pi-\pi^k||, \pi\in\mathbb{R}^T \}$
\IF{the NE is reached} 
\STATE each player $i\in\mathcal{N}$ sets $\Theta^{k+1}_i \gets (1-\rho^k).\Theta^k_i + \rho^k.\Theta_i^{k+1}$
\STATE Central operator sets $\pi^{k+1} \gets (1-\rho^k)\pi^k+\rho^k\pi^{k+1}$
\ENDIF
\STATE $k\gets k+1$
\ENDWHILE
\end{algorithmic}
\end{algorithm}

Convergence of Algorithm 2 is studied in Theorem \ref{algognep}.

\begin{thm}\label{algognep} Let the game $\mathcal{G}$ \eqref{GNEP} for [CP] be a GNEP with a nonempty set solution. If
\begin{enumerate}
    \item the regularization parameter $\tau$ satisfies:
        \begin{equation}
        \tau > \alpha.(N-1) + \sqrt{\alpha^2.(N-1)^2+4.N}
        \end{equation}
    \item $\rho$ is chosen such that $\rho \subset [R_m, R_M]$, with $0< R_m < R_M < 2$,
\end{enumerate}
then, any sequence $\{(\Theta^k,\pi^k)\}_{k=1}^{\infty}$ generated by PDA converges to a variational equilibrium of the GNEP.
\end{thm}
The monotonicity of $F$ ensures directly convergence \cite{AOSPF14}.
The [Net,VCG] cases are more sensitive. From Section \ref{Casestu}, we observe convergence with low inefficiency for
\[
\tau > 2\alpha (N-1) \underset{i\in\mathcal{N}}{\max}K_i + 2\sqrt{\alpha^2(N-1)^2\underset{i\in\mathcal{N}}{\max}K_i^2+N}.
\]
See Appendix \ref{appendixB}.

\section{Results and discussion}\label{Casestu}
\subsection{Use case}
 We study a REC  composed of 55 residential members connected behind the same MV/LV feeder. For the non-flexible loads, hourly electricity consumption profiles are extracted from the Pecan Street Project dataset \cite{Pecan}, and generated for whole days, with $T=24$. Battery Storage Systems are assigned to community members with a penetration level of 50\%. The initial battery state-of-charge is fixed at 50\%. Installed PV capacities vary between 0 and 10 kWc. The day ahead energy scheduling models are run for 20 days (10 days with high PV production - 10 days with low PV). We assume that the total daily energy demand remains constant, regardless of the schedule of flexible appliances (no load shedding). The prosumers may own different flexible devices: white goods (dishwasher, washing machine, clothes dryer, etc.), Electric Vehicles (EV) and Heat Pumps (HP). For simplicity and without loss of generality, the latter is considered as a fully flexible load.
We consider bi-hourly commodity tariffs: $\lambda_{imp}^t=0.08$ \euro/kWh, $\lambda_{exp}^t=0.02$ \euro/kWh, $\lambda_{iloc}^t=0.065$ \euro/kWh and $\lambda_{eloc}^t=0.032$ \euro/kWh between 21pm and 4am, and $\lambda_{imp}^t=0.16$ \euro/kWh, $\lambda_{exp}^t=0.04$ \euro/kWh, $\lambda_{iloc}^t=0.13$ \euro/kWh and $\lambda_{eloc}^t=0.05$ \euro/kWh elsewhere. We assume constant network tariffs: $\alpha=0.00109488$ \euro/kWh$^2$ and $\beta=0.1096737$ \euro/kW.

\subsection{Benchmark and Key Performance Indicators}
We simulate the use-case REC in the case of designs D1 and D2. We compare outcomes with a benchmark in which each user $i$ minimizes individually her own commodity and peak power cost (situation without community). All the convex quadratic models arising from the models are coded in Julia/JuMP and solved using Gurobi.
    
In each case, we compute the total REC costs (total costs are obtained in the individual benchmark by summing over the individual costs of each member first, and then adding the upstream grid costs). We also show individual costs (which are obtained directly when using the game theoretical models, or computed ex-post with the centralized optimization models). We report CPU solving times. We further calculate the following technical Key Performance Indicators (KPIs):
\begin{itemize}
\item \textbf{Self-Consumption Ratio (SCR):} ratio between the production consumed locally and the total production
\begin{equation} 1-\frac{\sum_{t\in\mathcal{T}}\sum_{i\in\mathcal{N}}\kappa^{t}_i}{\sum_{t\in\mathcal{T}}\sum_{i\in\mathcal{N}}g^t_i}, \text{ with } \kappa^t_i = 
\begin{cases}
l^{t-}_i & \text{if D1}\\
e^{ret,t}_i & \text{if D2.}
\end{cases}
\end{equation}
\item \textbf{Self-Sufficiency Ratio (SSR):} ratio between the load supplied locally and the total consumption
\begin{equation}  1-\frac{\sum_{t\in\mathcal{T}}\sum_{i\in\mathcal{N}}\kappa^{t}_i}{\sum_{t\in\mathcal{T}}\sum_{i\in\mathcal{N}}l^{t}_i+g^t_i}, \text{ with } \kappa^t_i=
\begin{cases}
l^{t+}_i & \text{if D1} \\
i^{ret,t}_i & \text{if D2.}
\end{cases}
\end{equation}
\item \textbf{Peak to Average Ratio}$^{+ (-)}$ \textbf{(PAR}$^{+(-)}$): ratio between the peak community consumption(+)/injection(-) and the average community consumption/injection
\begin{equation} \frac{T.\max_{t\in\mathcal{T}}\sum_{i\in\mathcal{N}}l^{t+(-)}_i}{\sum_{t\in\mathcal{T}}\sum_{i\in\mathcal{N}}l^{t+ (-)}_i}.
\end{equation}
\end{itemize}

\subsection{Results}
\subsubsection{Centralized formulations}
\begin{figure}
\centering
\includegraphics[width=0.9\textwidth]{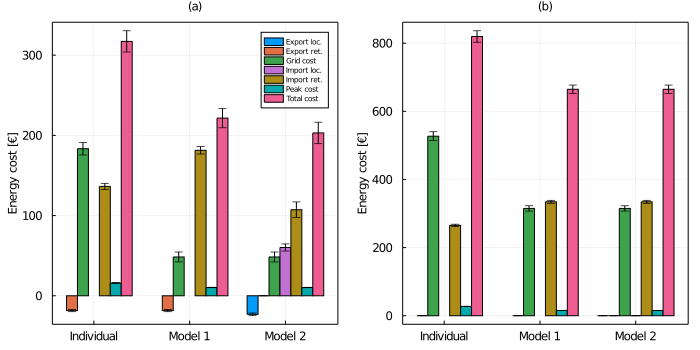}
\caption{Mean (and standard deviation) of the components of the total REC costs for the individual benchmark, D1 and D2, (a) for the high PV days and (b) the low PV days}\label{fig-2}
\end{figure}

Table \ref{tab-1} depicts the mean and standard deviations (between parentheses) of the total REC costs and technical KPIs, over the 10 high and low PV generation days, for the centralized models of section \ref{optiproblem}. The total REC cost components are shown in Fig. \ref{fig-2}. Simulation times reach 1s for design D1 with 6631 variables, and 2s for design D2 with 11911 variables.
 
\begin{table}[b]
\centering
\caption{Centralized results summary}\label{tab-1}
\small
\begin{tabular}{|l@{\hspace{0.2cm}}|c@{\hspace{0.2cm}}c@{\hspace{0.2cm}}c@{\hspace{0.2cm}}c@{\hspace{0.2cm}}c|}
\cline{2-6}
\multicolumn{1}{c|}{}
& REC Bill & PAR+ & PAR- & SCR [\%] & SSR [\%]\\ \hline
Day &   &   & High PV &   &  \\
Individual & 317.15 (13.24) & 2.37 (0.03) & 4.44 (0.28) & 73.08 (1.4) & 47.49 (0.78) \\
D1 & 221.55 (12.12) & 1.19 (0.03) & 2.45 (0.07) & 73.08 (1.4) & 47.49 (0.78) \\ 
D2 & 203.02 (13.42) & 1.19 (0.03) & 2.46 (0.07) & 100 (0) &  65 (2.3) \\ \hline
Day &   &   & Low PV &   &   \\
Individual & 819.55 (16.83) & 2.22 (0.01) & 4.8 (10.12) & 99.99 (0.03) & 1.49 (0.4)\\
D1 & 664.65 (12.42) & 1.21 (0.01) & 4.8 (10.12) & 99.99 (0.03) & 1.49 (0.4)\\ 
D2 & 664.65 (12.42) & 1.21 (0.006) & 4.8 (10.12) & 100 (0) & 1.5 (0.4) \\ \hline
\end{tabular}
\end{table}

Overall, we observe that design D2 leads to the lowest total cost, and to the best technical KPI values. The interest of operating as a community is particularly emphasized during high PV days. On average, total costs are 30.14\% (18.9\%) lower for high(low) PV days for community design D1 (see Table \ref{tab-1}). The gain mainly originates in lower upstream grid costs (and to a lesser extent individual peak costs), which is confirmed by the lower PAR values. Extra savings of 8.36\% are obtained with the community design D2 compared to D1, for high PV days. The excess of local production is exchanged inside the community at more advantageous prices than the retail markets prices, reducing commodity costs and the total REC bill. Design 2 also achieves higher community SCR and SSR values.\\

\subsubsection{Design D1 as a non-cooperative game}\label{stud1}
We first study the total REC costs and technical KPIs for the three cost-distribution methods [Net,VCG,CP] proposed in the noncooperative day-ahead scheduling problems \eqref{NEP}. Table \ref{tab-3} summarises the mean and standard deviations of the results. Simulation times reach between 480s and 780s for [Net, CP] and between 780s and 1200s for [VCG]. Note that if the initial values correspond to a social optimum, then the algorithm converges after 2 iterations in a few seconds for [Net,VCG].

\begin{table}
\centering
\caption{Decentralized model 1 results summary}\label{tab-3}
\small
\begin{tabular}{|l@{\hspace{0.2cm}}|c@{\hspace{0.2cm}}c@{\hspace{0.2cm}}c@{\hspace{0.2cm}}c@{\hspace{0.2cm}}c|}
\cline{2-6}
\multicolumn{1}{c|}{}
& REC Bill & PAR+ & PAR- & SCR [\%] & SSR [\%]\\ \hline
Day &   &   & High PV &   &  \\
CP & 224.85 (11.92) & 1.43 (0.02) & 2.42 (0.07)  & 73.08 (1.4) & 47.49 (0.78) \\
Net & 221.55 (12.12) & 1.19 (0.03) & 2.45 (0.07) & 73.08 (1.4) & 47.49 (0.78) \\ 
VCG & 221.55 (12.12) & 1.18 (0.03) & 2.45 (0.07) & 73.08 (1.4) & 47.49 (0.78) \\ \hline
Day &   &   & Low PV &   &   \\
CP & 669.54 (12.41) &  1.39 (0.003) & 4.8 (10.12) & 99.99 (0.03) & 1.49 (0.4)\\
Net & 664.65 (12.42) & 1.21 (0.004) & 4.8 (10.12) & 99.99 (0.03) & 1.49 (0.4) \\ 
VCG & 664.65 (12.42) & 1.21 (0.005) & 4.8 (10.12) & 99.99 (0.03) & 1.49 (0.4) \\ \hline
\end{tabular}
\end{table}

Results confirm Theorem \ref{optine}. Indeed, the optimal value of the total costs corresponds to the social optimum of Table \ref{tab-1}: 221.55\euro \- for [Net,VCG]. In addition, the standard deviations are identical. However, it is important to note that, while these cost distribution approaches tend to minimize the total REC costs, some users may experience lower profits, as it will be shown later. [CP] leads on the other hand to a sub-optimal solution. We quantify the inefficiency as $(\sum_{i\in\mathcal{N}}b_i^{\text{CP1}*}-C^{*}_{\mathcal{N}})/C^{*}_{\mathcal{N}}$, where $C^*_{\mathcal{N}}$ is the social optimum. The inefficiency remains however small: we obtain a mean inefficiency of 1.49\% (standard deviation of 0.27\%) for the high PV days, and 0.73\% (0.014\%) on the low PV days. In fact, the upstream grid contribution of \eqref{b1} is subject to strategy in this billing, explaining the greater inefficiency during high PV days. The PARs differ from the two other billings, whereas SCR and SSR remain the same.\\

\begin{figure}
\centering
\includegraphics[width=0.9\textwidth]{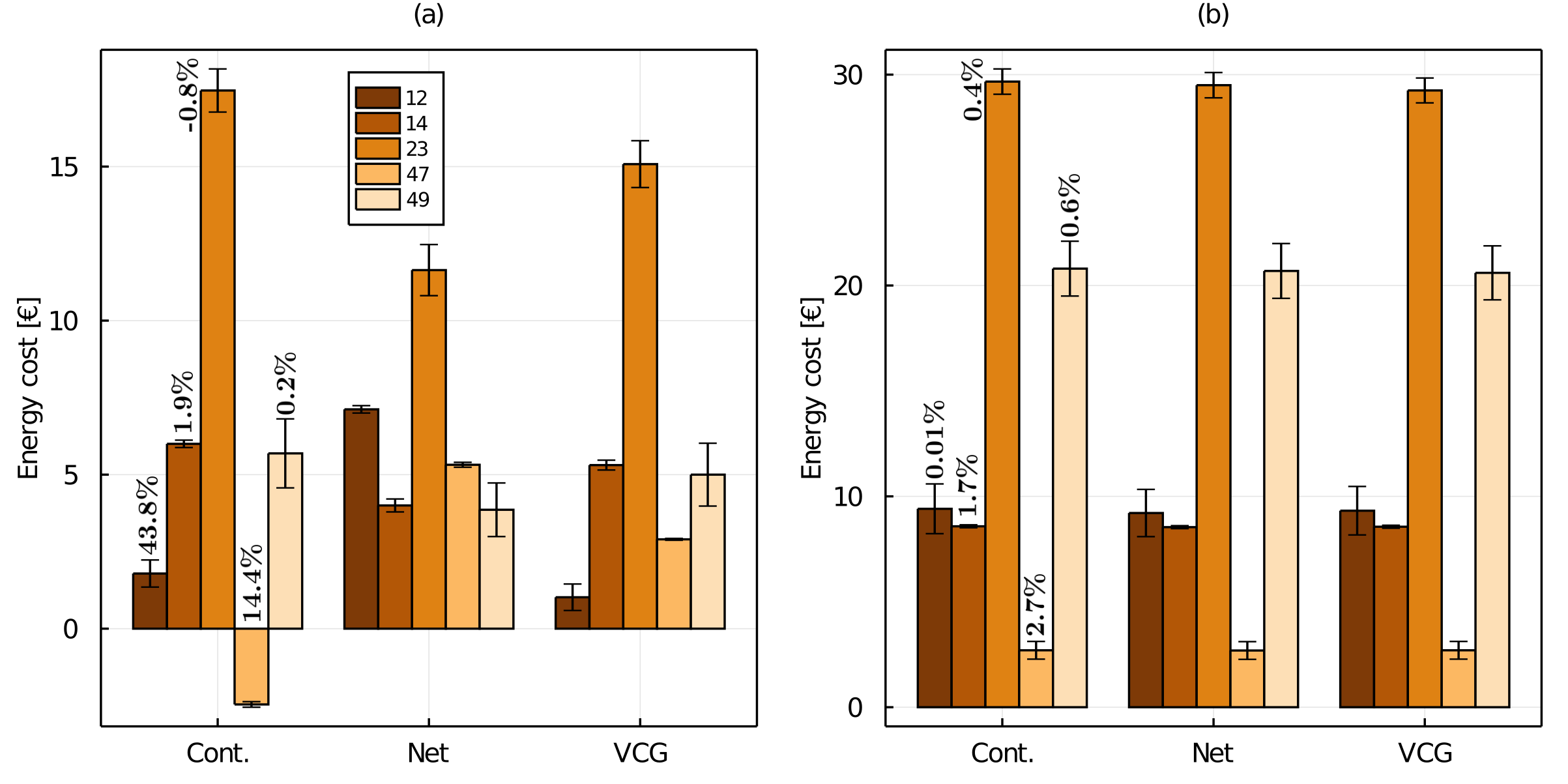}
\caption{Mean (and standard deviation) individual bills for the NEP \eqref{NEP} (a) on the high PV days and (b) on the low PV days. For each member, in the [CP] case, we superimpose the average percentage of change between the bill $b^{\text{CP}1}_i$ and the one obtained after ex-post allocation of the social optimum.}\label{fig-3}
\end{figure}

We then compare the individual member bills with those obtained in the centralized case. The latter are obtained ex post, i.e., the costs are distributed according to [Net,VCG,CP] after minimizing the REC bill \eqref{CNM}. Fig. \ref{fig-3} depicts the individual invoices of 5 selected users from the 55 members set, at the computed Nash equilibrium. Table \ref{tab-2} shows the characteristics of these members (electrical equipment, energy needs over the horizon, energy flexibility level computed as the ratio between available flexibility including battery storage and total consumption over a day). For each member, in the [CP] case, we superimpose on Fig. \ref{fig-3} the average percentage of change between the bill  $b^{\text{CP}1}_i$ and the one obtained after ex-post allocation of the social optimum (we do not show the changes in the case of [Net,VCG] since they are reasonably negligible). The change reaches 43.8\% and 14.4\% for members 12 and 47 respectively. However, their bill is rather low in absolute value, which explains the large relative change. Therefore, we conclude that, in the D1 case, the costs distribution among members remains very similar when solving the optimization problem \eqref{CNM} and the game \eqref{NEP}. 

\begin{table}[b]
\centering
\caption{End-users characteristics}\label{tab-2}
\small
\begin{tabular}{|c|c|c|c|c|}
\cline{1-5}
\multicolumn{1}{|c|}{}
& PV & ESS  & Total & Flexibility \\
& & (capacity, max. power) & consumption & level \\ \hline
User 12 & 9 kWc & (0 kWh, 0 kW) & 37.36 kWh & 0\%  \\
User 14 & 0 kWc & (0 kWh, 0 kW) & 33.48 kWh & 54.37\% \\
User 23 & 3 kWc & (0 kWh, 0 kW) & 116.03 kWh & 36.38\% \\ 
User 47 & 9 kWc & (14 kWh, 5 kW) & 11.84 kWh & 138.24\%  \\ 
User 49 & 8 kWc & (14 kWh, 5 kW) & 82.21 kWh &  78.1\% \\ \hline
\end{tabular}
\end{table}

Overall, the differences between cost distributions are more noticeable on days with high PV (Fig. \ref{fig-3}(a)), while they are very weakly discriminating for low PV days (Fig. \ref{fig-3}(b)). 

In [CP], we observe that members tend to be more remunerated if they consent to more flexibility, own PV production and have few consumption (member 47 benefits e.g. from negative mean prices). However, if they don't have storage and grant few flexibility as member 12 and 23, their outcomes tend to degrade compared to [VCG]. In brief, the continuous billing tends to give more strategy power to members and leads to a slight inefficiency.

The [Net] scheme does not take into account the individual users' impact on the total bill. In fact, the flexibility is not valued at the individual level, but shared among all members. It incentives members to minimize interactions with the network. Therefore, a prosumer who injects his surplus into the grid will have a bad outcome and contribute to increase the denominator in \eqref{net} (e.g, player 47). This method could be considered more advantageous for member profiles such as 14, 23 and 49, but less incentivizing for agent 12 without flexibility and storage, and for user 47 who have to resell its surplus. Hence, the net load proportional billing is efficient, tends to be more egalitarian, but seems to less incentivize flexibility.

The [VCG] scheme depends on the cost structure and the individual profiles. Some prosumers have a negative relative contribution, indicating in fact a positive impact of the member on the REC total costs (e.g., player 47). However, since we consider the absolute marginal costs, the distribution key is positive, leading to an increase in the denominator. Therefore, the users 12, 14, 23 and 49 have some reduction in their bill compared to the continuous case, but not necessarily as high as in the net scheme.\\

\subsubsection{Design D2 as a non-cooperative game}\label{stud2}
Table \ref{tab-4} summarises the mean and standard deviations of the results of the GNEP resolution. If the set of initial values correspond to a social optimum, simulation times reach 107s for [Net, VCG] and 149s for [CP].

\begin{table}
\centering
\caption{Decentralized model 2 results summary}\label{tab-4}
\small
\begin{tabular}{|l@{\hspace{0.2cm}}|c@{\hspace{0.2cm}}c@{\hspace{0.2cm}}c@{\hspace{0.2cm}}c@{\hspace{0.2cm}}c|}
\cline{2-6}
\multicolumn{1}{c|}{}
& REC Bill & PAR+ & PAR- & SCR [\%] & SSR [\%]\\ \hline
Day &   &   & High PV &   &  \\
CP & 203.83 (13.6) & 1.24 (0.076) & 2.45 (0.07) & 100 (0) & 65.02 (2.3) \\
Net & 203.04 (13.42) & 1.19 (0.032) & 2.46 (0.07) & 100 (0) & 65.02 (2.3) \\ 
VCG & 203.05 (13.42) & 1.19 (0.032) & 2.46 (0.07) & 100 (0) & 65.02 (2.3) \\ \hline
Day &   &   & Low PV &   &   \\
CP & 668.05 (13.13) & 1.36 (0.02) & 4.8 (10.12) & 100 (0) & 1.49 (0.4)\\
Net & 664.67 (12.42) & 1.21 (0.006) & 4.8 (10.12) & 100 (0) & 1.49 (0.4) \\ 
VCG & 664.68 (12.42) & 1.21 (0.006) & 4.8 (10.12) & 100 (0) & 1.49 (0.4) \\ \hline
\end{tabular}
\end{table}

We observe that the total bill does not correspond exactly to the social optimum for the [Net] and [VCG] billings. Theorem \ref{optignep} fails indeed to conclude an equivalence with global optimization \eqref{CM}, as $K_1\neq \ldots \neq K_N$. We know that the optimal solutions are generalized Nash equilibria, but they may not be variational equilibria. As expected, [CP] leads to a sub-optimal solution. The inefficiencies (i.e. deviations from the social optimum) remain however very small, especially for [Net,VCG]: Table \ref{tab-5} reports the mean (and standard deviation) of the inefficiency, calculated as $(\sum_{i\in\mathcal{N}}b_i^{*}-C^{*}_{\mathcal{N}})/C^{*}_{\mathcal{N}}$, where $C^*_{\mathcal{N}}$ is the social optimum.
We observe more particularly that the inefficiency of [CP] is lower than for [CP] in D1, whereas less robust (i.e, with a greater standard deviation) for days with high PV production. The [Net] method is the closest from efficiency for both PV levels, and is also very robust. For low PV days, the inefficiency of the [VCG] scheme \eqref{vcg} is negligible, which is not the case for high PV days. It is also less robust than the [Net] method.

\begin{table}
\centering
\caption{Inefficiency for design D2}\label{tab-5}
\small
\begin{tabular}{|l|c c c|}
\cline{2-4}
\multicolumn{1}{c|}{}
& CP [\%] & Net [\%]  & VCG [\%] \\ \hline
High PV &  0.4 (0.35) &  0.007 (0.0009)  & 0.014 (0.19)\\ \hline
Low PV & 0.51 (0.01) & 0.003 (0.0002) & 0.004 (0.08)\\ \hline
\end{tabular}
\end{table}

\begin{figure}
\centering
\includegraphics[width=0.9\textwidth]{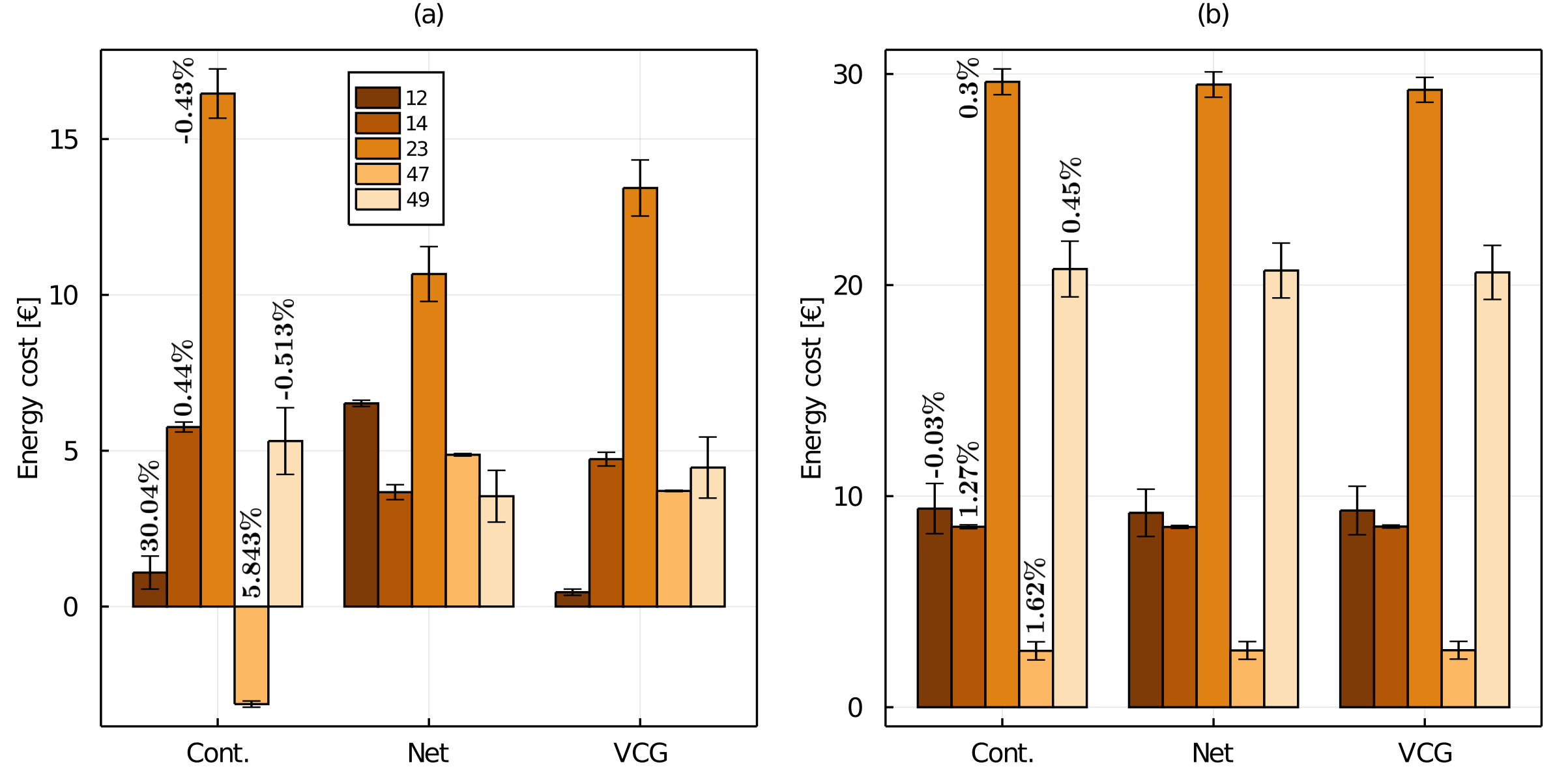}
\caption{Mean (and standard deviation) of members' costs of  GNEP \eqref{GNEP} (a) on high PV days and (b) on low PV days. For each member, in the [CP] case, we superimpose the average percentage of change between the bill $b^{\text{CP}2}_i$ and the one obtained after ex-post allocation of the social optimum.}\label{fig-4}
\end{figure}

Fig. \ref{fig-4} shows the individual invoices of the five users (Table \ref{tab-2}) at a computed variational equilibrium. We still note that selling local surplus energy inside the community allows for a reduction in the commodity costs. We observe that all members benefit from this saving in their individual invoice for [Net,CP]. For [CP], deviations between the individual bill obtained in decentralized model and after ex-post distribution of the social optimum are lower with design D2 (compared to D1 and Fig. \ref{fig-3}), wheras they are considered negligible for [Net] (max 0.02\%) and [VCG] (max 0.6\%). Other comments are similar than Section \ref{stud1}. 

\section{Conclusion}\label{concl}
This paper studied two market designs for the optimal day-ahead scheduling of energy exchanges within renewable energy communities. The first design (D1) implements a cooperative demand side management scheme inside a community where members objectives are coupled through grid tariffs, whereas the second design (D2) allows in addition the valuation of excess generation among community members and on the retail market. Both designs have been formulated as centralized optimization problems first, and as non cooperative games then. In the latter case, the existence and efficiency of the corresponding (Generalized) Nash Equilibria have been rigorously studied and proved, and distributed implementations of iterative solution algorithms for finding these equilibria have been proposed, with proofs of convergence. The models have been tested on a use-case made by 55 members with PV generation, storage and flexible appliances, and compared with a benchmark situation where members act individually (situation without community).  Compared to the benchmark, we showed that designs D1 and D2 saved 30.14\% and 36\% resp. on the whole REC bill. We studied three mechanisms in each design to share the REC total costs among members, and showed that the inefficiency of the decentralized models remains always very small.

This paper studies exclusively the short-term problem of energy assets dispatch. It would be interesting to extend the framework to other horizons, in particular to the investment in communities issue. In addition to cost reduction, other types of member preferences such as emission reduction and self-consumption could be considered in the models.

\appendix
\section{Nash Equilibrium Problem}\label{appendixA}

\subsection{Proof Theorem 4}
This section describes how the sequence generated by the proximal decomposition algorithm converges to a solution of the game $G=(\mathcal{N}, \Omega,(b_i)^N_{i=1})$ for [CP]. To do this, we exploit the equivalence between NEPs and VIs, following similar arguments than in \cite{AOSPF13b,SFPP14}. By \cite[Th. 17]{SFPP14}, we need to show that: (a) the mapping $F$ is monotone on $\Omega$; (b) the regularization parameter $\tau$ is large enough such that the $N \times N$ matrix $\Upsilon_{F,\tau}=\Upsilon_{F}+\tau I_N$ is a $P$-matrix (i.e, all principal minors are positive). The matrix $\Upsilon_{F,\tau}$ is related to the regularized VI$(\Omega,F+\tau.(I-y^k))$, with $\Upsilon_F$ defined by
\begin{align}
[\Upsilon_F]_{ij}&:=
\begin{cases}
\upsilon_i^{\min} & \text{if } i=j \\
-\upsilon_{ij}^{\max} & \text{if } i\neq j
\end{cases} \label{upsi} \\
\upsilon_i^{\min}&:= \underset{\Theta\in\Omega}{\min} \ \lambda_{\text{least}}(J_iF_i(\Theta))\\
\upsilon_{ij}^{\max}&:= \underset{\Theta\in\Omega}{\max} \ ||J_jF_i(\Theta)|| \label{vmax}
\end{align}
where $J_iF_i(\Theta)$ and $J_jF_i(\Theta)$ are partial Jacobian matrices of $F$, and $\lambda_{\text{least}}(.)$ is the smallest eigenvalue of the symmetric part\footnote{The symmetric part of a matrix $A$ is $\frac{1}{2}(A+A^\top$). If $A$ is symmetric, then $A=\frac{1}{2}(A+A^\top)$.} of argument matrix.

\begin{proof}[Proof of (a)] The mapping function $F=(\nabla_{\Theta_i}b_i)^N_{i=1}$ with $\Omega$ closed and convex, is monotone on $\Omega$ when $(\Theta-\Theta ')^\top.(F(\Theta)-F(\Theta ') \geqslant 0$, for all $\Theta, \Theta ' \in\Omega$. In the [CP] setting, the Jacobian matrix $JF$ is symmetric which allows us to assert $F$ is monotone if its symmetric Jacobian is a positive semidefinite matrix \cite{SFPP14}.
We can write the block elements of $JF(\Theta)$ as
\begin{align}
J_iF_i(\Theta)&:= \nabla^2_{\Theta_i,\Theta_i}b_i(\Theta)=\text{Diag}(H_{i}^1, \ldots, H_{i}^T, \mathbf{0}) \label{J_ii} \\ 
J_jF_i(\Theta)&:= \nabla^2_{\Theta_j,\Theta_i}b_i(\Theta)= \text{Diag}(J_jF_i^1, \ldots, J_jF_i^T, \mathbf{0}) \label{J_ij}
\end{align}
where $\mathbf{0}$ denoting zero matrix, $H_i^t$ and $J_jF_i^t$ are defined
\begin{align}
H_i^t&:= 
\begin{pmatrix}
0 & \cdots & 0 & 0 & 0 \\
\vdots & \ddots & \vdots & \vdots & \vdots \\
0 & \cdots & 0 & 0 & 0\\
0 & \cdots & 0 & 2\alpha & -2\alpha\\
0 & \cdots & 0 & -2\alpha & 2\alpha
\end{pmatrix}\\
J_jF_i^t&:=
\begin{pmatrix}
0 & \cdots & 0 & 0 & 0 \\
\vdots & \ddots & \vdots & \vdots & \vdots \\
0 & \cdots & 0 & 0 & 0\\
0 & \cdots & 0 & \alpha & -\alpha\\
0 & \cdots & 0 & -\alpha & \alpha
\end{pmatrix} \label{JFt}
\end{align}
After algebraic manipulations, it follows that the Jacobian matrix is a block diagonal matrix: $JF(\Theta):=\text{Diag}(\mathbf{0}, B^1, \ldots, B^T)$ with $B^t:=\alpha(D+E)$ for all $t\in\mathcal{T}$ given

\begin{align}
D:=
\begin{pmatrix}
1 & -1 & 0 & \cdots & 0 & 0\\
-1 & 1 & 0 & \cdots & 0 & 0\\
0 & 0 & \ddots &   & \vdots & \vdots\\
\vdots & \vdots & \vdots & \ddots & \vdots &\vdots \\
0 & 0 & 0 & \cdots & 1 & -1\\
0 & 0 & 0 & \cdots & -1 & 1
\end{pmatrix} \\
E:=((-1)^{m+v})_{m,v\in \{1,\ldots,2N\}}
\end{align}
 A block diagonal matrix is positive semidefinite if and only if each diagonal block is positive semidefinite. Furthermore, the sum of positive semidefinite matrices is positive semidefinite. Since this is trivial for the null matrix and as $\alpha>0$, if we show $D$ and $E$ are positive semidefinite then the proof is complete.

 In fact, $D$ is also a block diagonal matrix. The eigenvalues of $D$ are just the list of eigenvalues of each block. As each block is identical, we can easily calculate the eigenvalues 0 and 2 which are both nonnegatives. So, $D$ is positive semidefinite.

The rank of $E$ matrix is equal to 1, thus its kernel dimension is equal to $2N-1$ by Rank-nullity theorem. Therefore, 0 is an eigenvalue with multiplicity $2N-1$. All that remains is to find a vector that is not in the kernel, for example: 
$\begin{pmatrix}
    1 & -1 & 1 & -1 & \ldots & 1 & -1
\end{pmatrix}$,
to determine the last eigenvalue $2N$. The eigenvalues are nonnegatives, so the matrix $E$ is positive semidefinite.
\end{proof}

\begin{proof}[Proof of (b)]
We determine the value of $\tau$ such as $\Upsilon_{F,\tau}$ is a $P$-matrix.
The matrix \eqref{J_ii} correspond to the user $i$'s Hessian matrix. Because $b_i$ is convex on $\Omega_i$, its Hessian matrix is positive semidefinite. Therefore, we can state that $\upsilon_i^{\min}=0$ for all $i\in\mathcal{N}$. It remains to estimate the values of $\upsilon_{ij}^{\max}$ for all $i,j\in\mathcal{N}$, $i\neq j$. Considering $J_jF_i(\Theta) =\text{Diag}(J_jF_i^1, \ldots, J_jF_i^T, \mathbf{0})$ and $J_jF_i^t$ in \eqref{JFt}, we have: $\upsilon_{ij}^{\max}\leqslant2\alpha$.

Therefore, $\Upsilon_{F,\tau}$ is a $P$-matrix if the following condition is satisfied \cite[Prop. 7]{SFPP14} for all $i\in\mathcal{N}$, 
\begin{equation}\label{critnep}
\sum_{j\in\mathcal{N}\backslash\{i\}} \Big( \frac{\upsilon_{ij}^{\max}}{\upsilon_i^{\min}+\tau}\Big)\leqslant \sum_{j\in\mathcal{N}\backslash\{i\}} \frac{2\alpha}{\tau} \leqslant \frac{2\alpha(N-1)}{\tau} < 1.
\end{equation}
Consequently, any parameter $\tau$ such as:
\begin{equation}
\tau > 2\alpha(N-1)
\end{equation}
holds the criterion \eqref{critnep}. This completes the proof.
\end{proof}

\begin{rem}
The proof is less direct in the case of [Net,VCG]. Using similar reasoning we obtain a regularization parameter $\tau$ larger enough such that the $N$ square matrix $\Upsilon_{F,\tau}$ is a $P$-matrix
\begin{align}
\underset{\Theta\in\Omega}{\max} \ ||J_jF_i(\Theta)|| &\leqslant \underset{t\in\mathcal{T}}{\max} \Big( \underset{\Theta\in\Omega}{\max} \ \lambda_{\max}(J_jF_i^t)\Big) \notag \\
& \leqslant 4\alpha(N-1)\underset{i\in\mathcal{N}}{\max} \ K_i.
\end{align}
where $\lambda_{\max}(.)$ is the largest eigenvalue of the argument matrix.
However, the function $F$ is not monotone, but it is a $P_0$ function\footnote{$F$ is a P$_0$ function on $\Omega$ if for all pairs of distinct tuples $x,y \in \Omega$, an index $i$ exists such that $x_i\neq y_i$ and $(x_i-y_i)^\top (F_i(x)-F_i(y))\geq0$. If the inequality is strict, then $F$ is a P function.} on $\Omega$ which is convex and compact. Furthermore, the NE set is non-empty and compact by Theorem \ref{exine} and we noted convergence to a NE in our case-study framework.

Since we are dealing with a potential game, we know by Theorem \ref{optine}.1) and \cite[Lem. 2.1]{MS96} that solving $G=(\mathcal{N},\Omega,(b_i)_{i=1}^N)$ amounts to finding all equilibria of the game $G_P=(\mathcal{N},\Omega,(P)_{i=1}^N)$ with $P=f^1$. Because $\nabla P=\nabla f^1$ is monotone, we can obtain a Nash equilibrium via the PDA.

\end{rem}

\section{Generalized Nash Equilibrium Problem}\label{appendixB}
\subsection{Proof Theorem 8}
The convergence proof of the PDA with shared constraints for [CP] is also based on the relation between the extended NEP and VIs. The proof follows the lines of argument in \cite{AOSPF14,SPFP11}. Recalling Lemma \ref{lem2}, solving the extended NEP $\mathcal{G}_{ext}$ in (34) is the same as solving the VI$(\mathcal{Y},F_{ext})$, with $\mathcal{Y}:=(\prod_{i\in\mathcal{N}}\overline{\Omega}_i) \times \mathbb{R}^T$ and
\begin{equation}\label{Fext}
    F_{ext}(\Theta,\pi):=
    \begin{bmatrix}
        F(\Theta) + \pi^\top \nabla_{\Theta}h(\Theta) \\
        -h(\Theta)
    \end{bmatrix}
\end{equation}
Then, we solve a regularized sequence of VI$(\mathcal{Y}, F_{ext}+\tau(I-(y^k,\eta^k)))$ with $(y^k,\eta^k)$ in $\mathbb{R}^n \times \mathbb{R}^T$.
According to \cite[Th. 17]{SFPP14}, we need to show that: (a) the mapping $F_{ext}$ in \eqref{Fext} is monotone on $\mathcal{Y}$; (b) the regularization parameter $\tau$ is large enough such that the $N+1$ square matrix $\overline{\Upsilon}_{F,\tau}$ is a $P$-matrix
\begin{align}
    \overline{\Upsilon}_{F,\tau} &:=
    \begin{pmatrix}
        \Upsilon_F + \tau I_N & -\mu \\
        -\mu^{\top} & \tau
    \end{pmatrix}\\
    \mu &:= \Big(\max_{\Theta_i\in\overline{\Omega}_i} ||\nabla_{\Theta_i}h_i(\Theta_i)||_2\Big)_{i=1}^N 
\end{align}
with $\Upsilon_{F}$ defined in \eqref{upsi}-\eqref{vmax}.

\begin{proof}[Proof of (a)] If $F$ is monotone on $\prod_{i\in\mathcal{N}}\overline{\Omega}_i$, then $F_{ext}$ is so on $\mathcal{Y}$ \cite[Prop. 4.4]{SPFP11}. 
Since the matrices are similar to those obtained with model D1, we can directly conclude that $F$ is monotone.
\end{proof}

\begin{proof}[Proof of (b)] We determine the value of $\tau$ such as $\overline{\Upsilon}_{F,\tau}$ is a $P$-matrix.
In fact $\overline{\Upsilon}_{F,\tau}$ is a $Z$-matrix (i.e., all off-diagonal elements are non positive). We write $\overline{\Upsilon}_{F,\tau} \geqslant \Tilde{\Upsilon}_{F,\tau}$, where $\geqslant$ indicates component-wise $\geqslant$ and 
\begin{equation}
[\tilde{\Upsilon}_{F,\tau}]_{ij} := 
\begin{cases}
    \tau & \text{if } i=j \\
    -2\alpha & \text{if } i\neq j \text{ and } i,j \neq N+1 \\
     -2 & \text{otherwise}
\end{cases}
\end{equation}
If $\tilde{\Upsilon}_{F,\tau} $ is a $Z$ and $P$ matrix, then $\overline{\Upsilon}_{F,\tau}$ is $P$ matrix \cite[Thm. 3.11.10]{CPS09}.
We have that $\tilde{\Upsilon}_{F,\tau} $ is a $P$-matrix if and only if the spectral radius of the matrix $\Gamma_{F,\tau}$ is less than 1 \cite{SFPP14}:
\begin{equation}
   [ \Gamma_{F,\tau}]_{ij} :=
    \begin{cases}
        0 & \text{if } i=j \\
        \upsilon_{ij}^{\max}/ \tau & \text{if } i\neq j \text{ and } i,j \neq N+1\\
        \mu /\tau & \text{otherwise}
    \end{cases}
\end{equation}
where the spectral radius is the maximum of the absolute values of its eigenvalues. According to the Gershgorin circle theorem, every eigenvalues of $\Gamma_{F,\tau}$ is contained in at least one of the Gershgorin disks $D(0,R_i)$ with $R_i=\sum_{j\neq i}|[\Gamma_{F,\tau}]_{ij}|.$ Hence, $\overline{\Upsilon}_{F,\tau}$ is $P$ matrix if for some $\omega>0$, the conditions are verified
\begin{align}
    \tau&> 2\alpha(N-1)+2\omega\label{tau2} \\
    \tau&> \frac{2N}{\omega}.
\end{align}
The $\omega$ value minimizing $\tau$ is a solution of the second-degree equation
\begin{equation}
    2\alpha(N-1)+2\omega = \frac{2N}{\omega} \Leftrightarrow 2\omega^2 + 2\alpha(N-1)\omega - 2N =0.
\end{equation}
By incorporating this data in \eqref{tau2}, we have
\begin{equation}
    \tau > \alpha (N-1) + \sqrt{\alpha^2(N-1)^2+4N}.
\end{equation}
\end{proof}

\begin{rem}
The case of [Net,VCG] is more sensitive. We calculate a regularization parameter $\tau$ larger enough such that the $N+1$ square matrix $\overline{\Upsilon}_{F,\tau}$ is a $P$-matrix. Let $i,j\in\mathcal{N}$, if $i\neq j$, we have $[\tilde{\Upsilon}_{F,\tau}]_{ij}=-4\alpha \max_i K_i$ and so
\[
\tau > 2\alpha (N-1) \underset{i\in\mathcal{N}}{\max}K_i + 2\sqrt{\alpha^2(N-1)^2\underset{i\in\mathcal{N}}{\max}K_i^2+N}.
\]
However, the function $F_{ext}$ is not monotone, but it is a $P_0$ function on $\mathcal{Y}$ which is closed and convex. Furthermore, the VE set is non-empty and compact, and in our use-case we observe convergence with low inefficiency.

Even though $\mathcal{G}$ is a potential game, the structure of its set of strategies is non Cartesian, so that we can not use the same arguments than the NEP case. Nevertheless, we can apply the method of \cite{FPS11}, being aware that convergence to a VE is not guaranteed although the generated sequence is on $\mathcal{C}$ and each limits point is a GNE. In fact, there are no proofs or indications concerning the type of equilibrium (VE or not) one could get depending on the initial inputs. In [Net,VCG] case, there is no theoretical guarantee that VEs offer the cost optimal value, but we have Theorem \ref{optignep}.1).
\end{rem}

\end{document}